\documentclass[sigconf,nonacm]{acmart}

\usepackage{xcolor}
\usepackage{soul}
\usepackage{float}
\usepackage{graphicx}
\usepackage{placeins} 
\usepackage{tabularx}
\usepackage{afterpage}
\usepackage{algorithm}
\usepackage{algpseudocode}

\AtBeginDocument{%
  }

\begin{document}

\title{Modeling Object Attention in Mobile AR for Intrinsic Cognitive Security}

\author{Shane Dirksen}
\email{shanedirksen@ucsb.edu}
\affiliation{%
  \institution{University of California,\linebreak Santa Barbara}
  \city{Santa Barbara}
  \state{California}
  \country{USA}}

\author{Radha Kumaran}
\email{rkumaran@ucsb.edu}
\affiliation{%
  \institution{University of California,\linebreak Santa Barbara}
  \city{Santa Barbara}
  \state{California}
  \country{USA}}

\author{You-Jin Kim}
\email{yujnkm@tamu.edu}
\affiliation{%
  \institution{Texas A\&M University}
  \city{College Station}
  \state{Texas}
  \country{USA}}

\author{Yilin Wang}
\email{yilin_wang@ucsb.edu}
\affiliation{%
  \institution{University of California,\linebreak Santa Barbara}
  \city{Santa Barbara}
  \state{California}
  \country{USA}}

\author{Tobias Höllerer}
\email{holl@cs.ucsb.edu}
\affiliation{%
  \institution{University of California,\linebreak Santa Barbara}
  \city{Santa Barbara}
  \state{California}
  \country{USA}
}

\renewcommand{\shortauthors}{Dirksen et al.}

\begin{abstract}
We study attention in mobile Augmented Reality (AR) using {\em object recall} as a proxy outcome. We observe that the ability to recall an object (physical or virtual) that was encountered in a mobile AR experience 
depends on many possible impact factors and attributes, with some objects being readily recalled while others are not, and some people recalling objects overall much better or worse than others. This opens up a potential cognitive attack in which adversaries might create conditions that make an AR user not recall certain potentially mission-critical objects. 
We explore whether a calibrated predictor of object recall can help shield against such cognitive attacks. We pool data from four mobile AR studies (with a total of \(1{,}152\) object recall probes) and fit a Partial Least Squares Structural Equation Model (PLS-SEM) with formative Object, Scene, and User State composites predicting recall, also benchmarking against Random Forest and multilayer perceptron classifiers. PLS-SEM attains the best \(F_{1}\) score in three of four studies. Additionally, path estimates identify lighting, augmentation density, AR registration stability, cognitive load, and AR familiarity as primary drivers. The model outputs per-object recall probabilities that can drive interface adjustments when predicted recall falls. Overall, PLS-SEM provides competitive accuracy with interpretable levers for design and evaluation in mobile AR.

\smallskip
\textit{ This is a preprint version of this article. The final version of this paper can be found in the Proceedings of the ACM MobiHoc 2025. For citation, please refer to the published version.}
\textit{This work was initially made available on the author's personal website [yujnkm.com] in September 2025, and was subsequently uploaded to arXiv for broader accessibility.}

\end{abstract}




\maketitle

\section{Introduction}
Consider a Forward Observer (FO) scanning for potential enemy positions, such as a small outpost at the edge of a field. An adversary, aware of the AR system’s capabilities, could distract the FO by launching a flare away from the outpost to pull gaze. They could increase visual clutter with a swarm of drones to raise cognitive load. They could also trick the system into showing overlays by placing decoy panels or dummy equipment that the software mistakes for valid targets. By partially concealing the outpost with smoke, the adversary could further reduce its visibility and make it less likely to draw attention. The risk is not that the outpost goes unseen, but that attention is diminished at critical times. To counter this, the MR system can filter excess cues, reduce clutter, or briefly highlight the outpost with a virtual overlay to keep it in focus. It can also ease cognitive load by simplifying its display at critical times, limiting nonessential audio cues, or adding simple visual guidance that directs attention back to key objects.

Mixed reality (MR) and Augmented Reality integrate real and virtual environments in real time. Adversaries can exploit this close coupling between users and MR systems by targeting cognitive processes through techniques such as flooding the scene with information, placing real objects to clutter displays, injecting virtual data to divert attention, or triggering false alarms \cite{teymourian2025sok}. Such attacks have been shown to induce cybersickness, confusion, anxiety, emotional shifts, and loss of trust \cite{cheng2023exploring, lebeck2017securing, yang2024inception}. It is desirable to prepare MR and AR systems to shield against such attacks via cognitive security \cite{huang2023cognitive} efforts.

In this work, we explore a particular cognitive security pattern, concerned with {\em attention attacks}, such as salient distractions interfering with target search and awareness or attackers finding ways to instill cognitive load by creating spurious activity that they know will have to be monitored by the AR system and human observer. FOs and Joint Terminal Attack Controllers (JTACs) provide a mission example where we seek to protect against cognitive attacks. They use mixed reality headsets to stake out observation points, maintain situational awareness, and mark targets during close air support. Because AR displays compete for attention, an adversary can manipulate the system by introducing distractions, decoys, or overlays that divert focus at critical moments. Intrinsic Cognitive Security (ICS) treats this as a question of risk and seeks probabilistic guarantees on human performance. In our work, we measure attention using object recall. Inattentional blindness studies show that unattended items are rarely recalled \cite{simons_gorillas_1999}. With that proxy in place, we explore how adversaries might disrupt attention to mission-critical objects in AR, and how mitigation could counter these effects.

\begin{figure}[!t]
  \centering
  \includegraphics[width=\columnwidth]{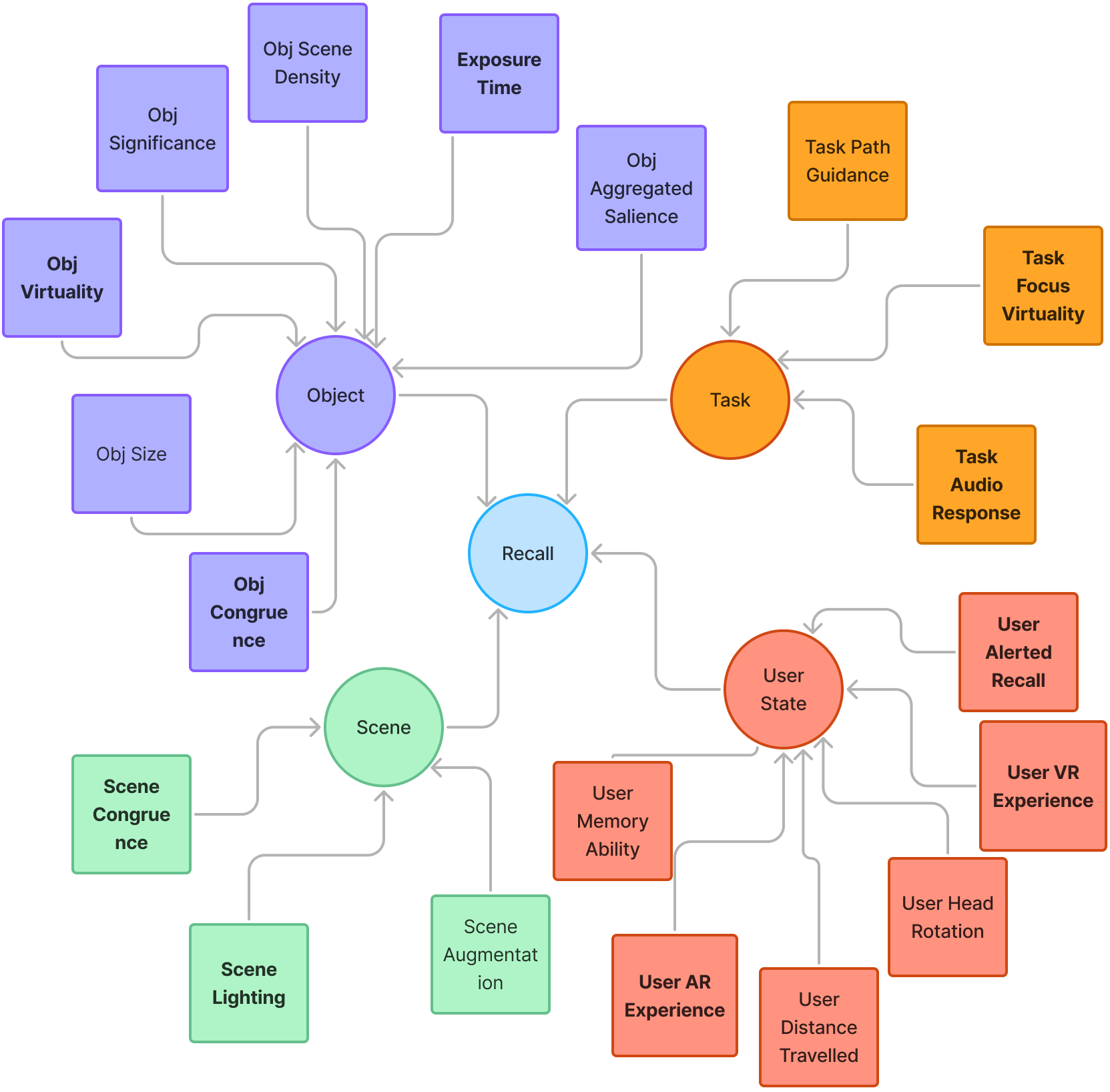}
  \caption{Our overall theorized model. Bold variables are already present in our model. Future data collection will include all listed variables.}
  \Description{Theorized model diagram; bold entries are already implemented.}
  \label{fig:theorized_model}
\end{figure}

We analyze object recall with PLS-SEM and benchmark it against two machine-learning baselines: Random Forests (RF) and a Multilayer Perceptron (MLP).  This study unifies four prior Augmented Reality datasets (with a total of \(1{,}152\) object recall events) that vary in scenes, objects, users, and tasks. Our complete theorized model specifies four latent constructs: Task, Object, Scene, and User State as latent predictors of Object Recall (Figure \ref{fig:theorized_model}). Because of current data limitations, we omit Task and adjust some indicators, including reassignments to avoid latent variables with only two indicators, so the present analysis uses Object, Scene, and User State. The path coefficients expose practical levers for attention, including scene attributes (e.g., lighting, virtual/physical congruence) object attributes (e.g., virtuality, object congruence with the scene), and user attributes (e.g., AR and VR familiarity). Taken together, this yields two avenues toward our goal: achieving competitive predictive accuracy, and identifying the conditions under which baseline performance holds or can be restored when attacks degrade it.

We make three contributions: (1) modeling attention via an object-recall proxy in realistic mobile AR scenarios; (2) a comparative evaluation of PLS-SEM against Random Forests and an MLP using identical cross-validation; and (3) a model that informs mitigations that help sustain attention during cognitive attacks.

\FloatBarrier

\section{Background}

\subsection{Cognitive Attacks in Mixed Reality}

Recent literature at the intersection of security and human-computer interaction, with a specific focus on MR/VR/AR systems, has explored various potential attacks on human sensing in MR experiences, including perception manipulation \cite{tseng2022dark,cheng2023exploring,cheng2024user,yang2024inception,casey2019immersive,azim2025your}, UI attacks \cite{cheng2024user,lee2025illusion,yang2024inception}, deception attacks \cite{teymourian2025sok}, and visual hindrance \cite{lebeck2017securing}. The attention phenomena and attacks modeled in this work align broadly with this latter category, but also have contact points with UI attacks. Overall, our focus is on modeling the attention-relevant concept {\em object recall}, so that the impact factors of increased recall can be used to counteract potential attention attacks.

\subsection{Modeling Attention}
Computational models treat attention as a competition between bottom-up salience and top-down task goals \cite{itti201423}. A standard baseline builds a saliency map from multi-scale contrasts in color, intensity, and orientation to predict likely fixations, showing how conspicuous items can pull gaze even when the user is goal directed \cite{itti_model_1998}. Even subtle variations in low-level visual features can shift attentional timing \cite{duinkharjav_image_2022}, indicating how bottom-up factors shape the dynamics of this competition. We view MR interfaces through that lens: attacks increase the bottom-up salience of distractors placed near or over mission targets, while mitigations strengthen the top-down priority on mission cues. We use object recall as the behavioral readout to test whether task-relevant items win this competition.

\subsection{Structural Equation Modeling (SEM)}
Jöreskog’s 1970 LISREL paper framed SEM as a way to encode a theory with unobserved constructs connected by hypothesized paths \cite{joreskog_general_1970}. Each construct is anchored by observed variables that measure it, and the links among constructs capture the causal story. Parameters are estimated so the model-implied covariances approximate the sample covariances, commonly via maximum likelihood. SEM asks whether the structure is plausible and whether the path estimates support the theory, with emphasis on explanation rather than out-of-sample prediction.

\subsection{Partial Least Squares SEM}
Wold introduced partial least squares path modeling in 1982 \cite{nijkamp_systems_1985}. Like SEM, it represents latent constructs and their relationships with observed indicators, but instead of fitting covariances to test a theory, PLS-SEM builds composites to maximize explained variance. PLS‐SEM \cite{sarstedt2021partial} iteratively applies ordinary least squares (OLS) to create latent composite scores that maximize the explained variance of dependent constructs (high $R^{2}$). It makes few distributional assumptions, handles small samples, and easily models formative composites where indicators define the construct. Model quality is judged by predictive metrics and the size of path coefficients, not by global fit tests. These traits make PLS‐SEM a practical fit for our research.

\section{Object Recall Data}
To populate our model, we are using data from four previous mobile augmented reality studies \cite{kim_investigating_2022,kumaran_impact_2023,kim_go_2025,kumaran2025scene}
that each explored aspects of object search in augmented outdoor (studies 1 and 2) and indoor (studies 3 and 4) environments. In all four studies, participants completed object recall tasks in which they were asked (either during the trial or post trial) whether they had encountered specific present objects (e.g., a fire hydrant) while completing the respective primary task, and, depending on the study, to classify the recalled object as physical or virtual. We define \textit{object recall} as the binary outcome of whether a participant correctly reported having encountered a given present object.

\subsection{Prior Studies}\label{subsec:range}

\begin{table}[t]
\centering
\caption{Constructs in Fig.~\ref{fig:theorized_model} and their indicators. \textbf{Bold} = currently implemented.}
\label{tab:fig1_indicators}
\footnotesize
\setlength{\tabcolsep}{3.5pt}
\begin{tabularx}{\columnwidth}{@{}l>{\raggedright\arraybackslash}X@{}}
\toprule
\textbf{Latent Construct} & \textbf{Indicators} \\
\midrule
Object &
\textbf{Exposure Time}; Object Aggregated Salience; \textbf{Object Congruence}; Object Scene Density; Object Significance; Object Size; \textbf{Object Virtuality} \\
Task &
\textbf{Task Audio Response}; \textbf{Task Focus Virtuality}; Task Path Guidance \\
User State &
\textbf{User Alerted Recall}; \textbf{User AR Experience}; User Distance Traveled; User Head Rotation; User Memory Ability; \textbf{User VR Experience} \\
Scene &
Scene Augmentation; \textbf{Scene Congruence}; \textbf{Scene Lighting} \\
\bottomrule
\end{tabularx}
\end{table}

\afterpage{%
\begin{table*}[!t]
\centering
\caption{Factors varied per study: Y = varied, N = not varied, C = captured (factor was manipulated in the study but did not vary at time of recall).}
\label{tab:object-recall-factors}
\resizebox{\textwidth}{!}{
\begin{tabular}{l cccc cc ccc ccc ccc}
\toprule
& \multicolumn{4}{c}{Object} & \multicolumn{2}{c}{Object/User} & \multicolumn{3}{c}{User} & \multicolumn{3}{c}{Task} & \multicolumn{3}{c}{Scene} \\
\cmidrule(lr){2-5}\cmidrule(lr){6-7}\cmidrule(lr){8-10}\cmidrule(lr){11-13}\cmidrule(lr){14-16}
Varied in: & Signif. & Size & Virtuality & Congruence & Agg. Salience & Exposure & Alerted Recall & AR Exp. & Memory & Focus Virt. & Audio Task & Path Guidance & Lighting & Congruence & Augmentation \\
\midrule
Study I (Lighting)                  & Y & Y & Y & C & Y & Y & C & Y & N & C & C & N & Y & C & C \\
Study II (Navigation Aids)          & Y & Y & Y & C & Y & Y & C & Y & N & C & C & N & Y & C & C \\
Study III (Clutter)                 & N & Y & Y & C & Y & Y & C & Y & N & C & N & C & C & C & C \\
Study IV (Adaptive Navigation Aids) & N & Y & C & Y & Y & Y & C & Y & Y & C & N & N & C & C & C \\
\bottomrule
\end{tabular}}
\end{table*}
}

The following provides a brief summary of the focus of each study and the details on each respective object recall task: 

\paragraph{Study I: Outdoor treasure hunt for virtual gems under different lighting conditions}
Forty-eight participants wearing HoloLens 2 searched a courtyard for green virtual gems and classified each while walking. Lighting (evening ambient vs. night) and cognitive load (gem task alone vs. with an auditory target-detection stream) were systematically manipulated; gems were placed free-floating, behind physical objects, or behind virtual objects. The researchers recorded head/gaze/position, walking paths, button responses, and then queried memory for encountered objects \cite{kim_investigating_2022}.

\paragraph{Study II: Outdoor treasure hunt for virtual gems with different AR navigation aids}
Twenty-four participants wearing HoloLens 2 searched a wide-area environment for 24 virtual gems and classified each by orientation (vertical/horizontal) and texture (rough/smooth) while walking. Conditions were within-subjects: in-world arrows, on-screen radar, and an on-screen horizontal compass. During search, participants also performed an audio target-detection task; afterward they completed an object-recall test classifying listed items. The researchers recorded head position/orientation, eye-gaze, movement, and responses for analysis \cite{kumaran_impact_2023}.

\paragraph{Study III: Indoor treasure hunt for virtual and physical gems with different scene augmentation density and controlled path guidance}
Twenty-four adults wearing HoloLens 2 walked an L-shaped indoor course (208 m²), searching for 12 gems per trial (6 physical, 6 virtual) and classifying each as marked vs. unmarked. Trials crossed augmentation density (low/high) with path guidance (spotlight ring present/absent); in guided trials a green ring set the path at 0.92 m/s. The researchers logged head rotation, distance, detections, and discrimination; after eight trials, participants completed a surprise object-recall test for six goal-irrelevant items (3 physical, 3 virtual) and reported noticing a highly salient “Godzilla” \cite{kim_go_2025}.

\paragraph{Study IV: Indoor treasure hunt for virtual gems with user choice of AR navigation aid and varied aid registration stability}
Twenty-four adults wearing HoloLens 2 searched an L-shaped hallway (208 m²) for 12 gems per trial (121 s), classifying shape. Baselines used no aid, arrows, or radar; in Mixed blocks participants could toggle world-locked arrows and an on-screen radar. Arrow reliability was manipulated (none, Mild latency, Severe with intermittent displacement). After each trial, participants completed an object-recall test with congruent vs. incongruent items \cite{kumaran2025scene}.
\\

For all studies, participant behavior was recorded and later reviewed via playback software that reconstructed the participant’s view of the scene (and eye gaze, when available), including all objects that appeared in the recall quizzes. Through user study playback exploration, we can procure future impact factors such as object clutter within the scene, object occlusion, user distance from object, dwell time, gaze hits, computational saliency scores, etc.

\subsection{Range of Participant and Object Performance}\label{subsec:range}

\textbf{Participants.}
Here we summarize recall aggregated across all present objects for each participant. In terms of participant performance on the recall tasks, the target variable \textit{recall}'s spread is large in every dataset. Study~1 ranges from 1.00 (9/9) down to 0.44 (4/9), range \(=0.56\). Study~2 ranges from 1.00 (6/6) to 0.33 (2/6), range \(=0.67\). Study~3 ranges from 1.00 (6/6) to 0.17 (1/6), range \(=0.83\) (largest). Study~4 ranges from 0.78 (14/18) to 0.06 (1/18), range \(=0.72\). Ceiling performance is common in Studies~1–3; Study~4 shows a lower ceiling and a heavy lower tail. While some of this spread is due to differences in user background, this also indicates an opportunity for attacks on participant parameters such as cognitive load and focus. 

\textbf{Objects.}
Object recall also spans a wide range. 
In Study~1, several ``twin'' items (items that occurred in the scene as both physical objects as well as virtual digital twin versions, see \cite{kim_investigating_2022}) are at 1.00 while the lowest performing object, a physical fire hydrant, is 0.60 (range \(=0.40\)). 
Study~2 has many twin/virtual items at 1.00, but two physical items (billboard, wagon) at 0.17 (range \(=0.83\), largest).
Study~3 tops at 0.88 (physical umbrella, virtual hammock) with a virtual coconut at 0.38 (range \(=0.50\)).
Study~4 tops at 0.67 (virtual arch) with small virtual items at the floor (camera, stool at 0.08; range \(=0.58\)).
Virtuality alone does not determine recall: both physical and virtual items appear at the top and bottom, demonstrating that item identity and context matter.

Studies with more items per participant show lower top-end recall and lower minima. Study~4 (18 objects) has a 0.78 ceiling and a 0.06 floor, while Studies~1–3 (6–9 objects) hit 1.00 for many participants. This pattern is consistent with memory load and supports including exposure and task-load indicators. The tails are wide by participant \emph{and} by object. An attacker can pick low-recall items (e.g., mundane physical fixtures outdoors, small virtual props indoors) and amplify pressure by raising augmentation density, destabilizing guidance, or degrading lighting. This motivates per-object scoring and targeted mitigation rather than uniform treatments.

\section{Methodology}

Our PLS-SEM model predicts object recall, a binary outcome variable indicating whether participants successfully remembered encountering specific objects during mixed reality tasks (single-indicator reflective). Predictors are three formative composites: \textit{Object} (virtuality: twin/virtual vs.\ physical; object–scene congruence), \textit{Scene} (lighting, scene congruence, normalized exposure time), and \textit{User State} (task focus, alerted-recall, audio task, AR/VR familiarity). The structural model includes direct paths from \textit{Object}, \textit{Scene}, and \textit{User State} to \textit{Object Recall}.

\begin{algorithm}[H]
\caption{SEMinR PLS-SEM estimation}
\label{alg:plssem}
\begin{algorithmic}[1]
\Require Indicator data matrix $X$, grouped into blocks $X_j$ for each construct $j$; structural model matrix $B$
\State Standardize all indicators in $X$
\State Give each indicator an initial, equal weight $w_{jk}$
\Repeat
  \State \textbf{Inner step (structural model):} 
    For each construct $j$, compute a temporary score 
    $\tilde z_j$ by combining the scores $z_\ell$ of connected constructs $\ell$ according to $B$
  \State \textbf{Outer step (measurement model):}
    \If{construct $j$ is reflective}
      \State Update weights $w_{jk} \gets \mathrm{cor}(x_{jk}, \tilde z_j)$
    \Else
      \State Update weights $w_j \gets (X_j^\top X_j)^{-1} X_j^\top \tilde z_j$
    \EndIf
  \State Normalize weights $w_j$ and update construct scores $z_j \gets X_j w_j$
\Until{the weights $w_j$ and scores $z_j$ converge}
\State Estimate final path coefficients $\beta$ for structural links using OLS regressions of $z_j$ on its predictors
\State Report $R^2_j$ for each dependent construct
\State Compute loadings $\lambda_{jk} = \mathrm{cor}(x_{jk}, z_j)$ (reflective) or inspect final weights $w_j$ (formative) and assess reliability/validity
\end{algorithmic}
\end{algorithm}

We estimate the PLS-SEM in R using the \texttt{seminr} package \cite{ray_seminr_2025}. Indicators are encoded, standardized within study, and incomplete rows are removed. Estimation follows the standard procedure outlined in Algorithm~\ref{alg:plssem}, with outer weights learned for formative blocks and the path-weighting scheme applied to the inner model. For prediction, we use $k$-fold cross-validation consistent with the machine learning baselines: in each fold the model is re-estimated on the training split, construct scores for the test split are formed using the trained outer weights, and recall probabilities are generated from the estimated structural paths.

To evaluate predictive performance across the different models, we compared our PLS model to two distinct machine learning approaches and assessed their classification accuracy using standard binary metrics: accuracy, precision, recall, and $F_1$. The PLS model was evaluated with 10-fold cross-validation, generating out-of-sample predictions that were thresholded at 0.5 against actual recall outcomes. The random forest used 500 decision trees with the square root of the number of features as the number of variables randomly sampled at each split. The multilayer perceptron was configured with a single hidden layer containing 10 neurons, L2 regularization with a decay parameter of 0.1, linear output activation, standardized inputs, and up to 1000 training iterations. All three approaches used identical cross-validation folds and the same thresholding procedure to enable direct performance comparisons.

\section{Results}

\subsection{Model Results}

\begin{figure*}[t]
  \centering
  \includegraphics[width=0.8\textwidth]{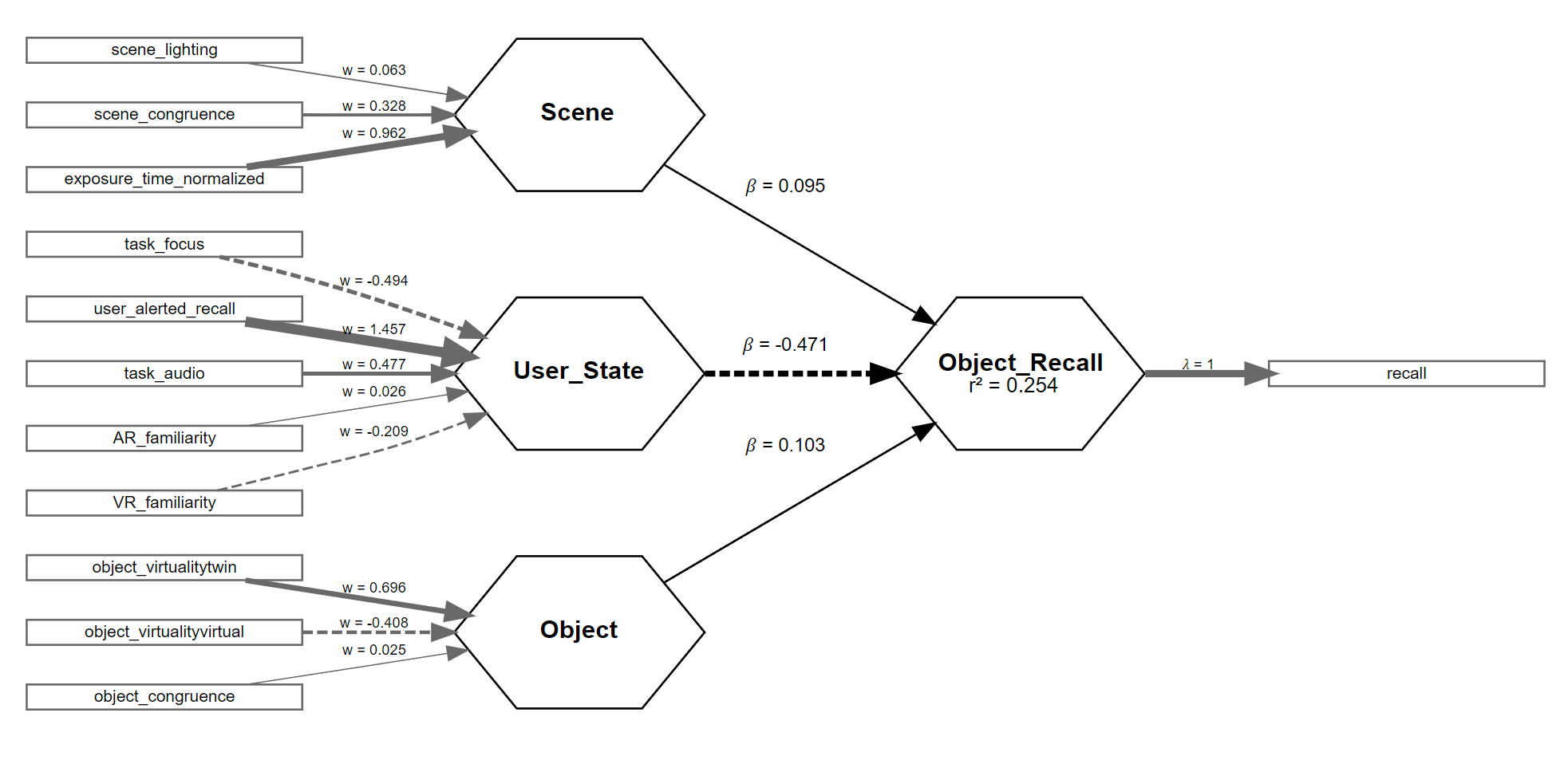} 
  \caption{The resulting PLS-SEM model from combining studies 1 to 4. 
  Rectangles denote observed indicators, hexagons denote latent constructs. 
  Arrows from indicators to constructs represent measurement paths with outer weights $w$, where larger $w$ means the indicator contributes more strongly to the construct. 
  Arrows between constructs represent structural paths with coefficients $\beta$; dashed arrows indicate negative paths. 
  $\lambda$ marks the loading of the single-indicator reflective construct. 
  $r^2$ shows the proportion of variance explained in the dependent construct.}
  \Description{Structural equation model combining studies 1 to 4.}
  \label{fig:sem_model}
\end{figure*}

Figure~\ref{fig:sem_model} explains $r^2=0.254$ for Object Recall. The largest structural path is User\_State $\rightarrow$ Object\_Recall ($\beta=-0.471$). Object ($\beta=0.103$) and Scene ($\beta=0.095$) are small and positive. Within Scene, exposure\_time\_normalized has the highest weight ($w\approx0.962$), with scene\_congruence ($w\approx0.328$) and scene\_lighting ($w\approx0.063$) smaller. Within User\_State, user\_alerted\_recall is the dominant indicator ($w\approx1.457$); task\_focus loads negatively ($w\approx-0.494$); task\_audio is modest ($w\approx0.477$; a second small loading $w\approx0.026$); AR\_familiarity is small and negative ($w\approx-0.209$). Within Object, object\_virtualitytwin is positive ($w\approx0.696$), object\_virtualityvirtual is negative ($w\approx-0.408$), and object\_congruence is near zero ($w\approx0.025$). Indicator weights are comparable only within a construct.

The large weight on \texttt{user\_alerted\_recall} reflects a between-study covariate: it is False in Studies 1–3 and True in Study 4. Study 4 also has lower recall (63\% failures). In the pooled model this indicator tags Study 4 and gets a large formative weight; with a negative \textit{User\_State} $\rightarrow$ \textit{Object\_Recall} path, a higher \textit{User\_State} score predicts lower recall. Per-study fits confirm no within-study effect of this indicator.

\subsection{Quantitative Results}

SEM and RF tie for best \(F_{1}\) (0.941, accuracy 0.889). In Study 2, SEM is highest (\(F_{1}=0.863\)). In Study 3, SEM is again highest (\(F_{1}=0.785\)); MLP shows 0.960 accuracy but only \(F_{1}=0.705\), consistent with a majority-class bias. Study 4 is the most difficult: accuracies are 0.583--0.648 and all \(F_{1}\) scores are low; MLP leads with \(F_{1}=0.444\). In the combined category, MLP has the top \(F_{1}\) (0.816); however, SEM is not far off (.803).

\subsection{PLS-SEM Range of Participant and Object Performance}\label{subsec:range}

\textbf{Participants.}
In Studies~1--3, participants with perfect actual recall also have SEM accuracy \(=1.00\), so the model preserves the ceiling cases (e.g., S1\_P15: \(9/9\) actual, SEM \(1.00\)). The lowest actual in Study~3 (0.17; \(1/6\)) also shows very low SEM accuracy (0.17), indicating misses concentrated in that extreme tail (e.g., S3\_P20: \(1/6\) actual, SEM \(0.17\)). Study~4 flips: participants with very low actual recall (0.06--0.22) have high SEM accuracy (e.g., S4\_P16: \(1/18\) actual, SEM \(0.94\)), while high-recall participants (0.72--0.78) fall into the SEM bottom tail (0.22--0.28; e.g., S4\_P3: \(14/18\) actual, SEM \(0.22\)).

\textbf{Objects.}
Twin items that were perfectly recalled (Studies~1–2) also yield SEM accuracy \(=1.00\), again matching the ceiling. Small virtuals in Study~4 with very low actual recall (camera 0.08, rug 0.13) have high SEM accuracy (0.83, 0.88), meaning the model predicts non-recall correctly for most trials. By contrast, several low-recall outdoor physicals in Study~2 (billboard, wagon at 0.17 actual) show SEM accuracy of 0.17, a strong mismatch with a majority baseline—here the model predicts the wrong class most of the time. Study~4 also contains objects where SEM underpredicts recall (e.g., arch: actual 0.67, SEM accuracy 0.33), consistent with that study’s overall difficulty and the model’s conservative predictions there.

Agreement is strongest at the extremes when the ground truth is near 0 or 1. Disagreement concentrates in Study~4 and in a subset of outdoor physical items in Study~2, where SEM either leans toward non-recall (Study~4) or incorrectly predicts recall (Study~2). This mirrors the earlier range analysis: ceiling cases are easy; the widest tails by study and object are where SEM accuracy is most variable.

\begin{table}[H]
\caption{Accuracy and $F_1$ on exposure–time datasets.}
\label{tab:perf}
\scriptsize
\setlength{\tabcolsep}{3.6pt}
\begin{tabular*}{\columnwidth}{@{\extracolsep{\fill}}lrrrrrrrc}
\toprule
 & \multicolumn{1}{c}{$N$} &
 \multicolumn{2}{c}{SEM} &
 \multicolumn{2}{c}{RF} &
 \multicolumn{2}{c}{MLP} &
 Best\\
\cmidrule{3-4}\cmidrule{5-6}\cmidrule{7-8}
Study & & Acc & $F_1$ & Acc & $F_1$ & Acc & $F_1$ & (by $F_1$)\\
\midrule
1 & 432  & 0.889 & \textbf{0.941} & 0.889 & \textbf{0.941} & 0.875 & 0.930 & SEM/RF\\
2  & 144  & 0.806 & \textbf{0.863} & 0.778 & 0.858 & 0.771 & 0.842 & SEM\\
3  & 144  & 0.646 & \textbf{0.785} & 0.681 & 0.698 & 0.960 & 0.705 & SEM\\
4  & 432  & 0.648 & 0.309 & 0.616 & 0.297 & 0.583 & \textbf{0.444} & MLP\\
\cmidrule{1-9}
Combined & 1152 & 0.748 & 0.803 & 0.761 & 0.814 & 0.757 & \textbf{0.816} & MLP\\
\bottomrule
\end{tabular*}
\end{table}

\section{Discussion}

We asked whether an interpretable predictor of object recall can act as an ICS control signal. The pooled PLS-SEM supports this: SEM achieves the highest $F_{1}$ in three of four studies and is close overall (Table~\ref{tab:perf}), which matters under class imbalance.

Objects show similarly broad variation: several “twin” items, which had more opportunities to be observed by participants and had increased salience because of their duplication across the physical and virtual realms, are at 1.00 while some physicals fall to the bottom tail when outdoors. Indoors, however, virtuality alone does not sort the winners and losers---identity and context matter. More items per participant coincide with lower maxima and minima (18 in Study~4 vs. 6–9 in Studies~1-2), consistent with increased memory load.

Our work points at several possible mitigations shielding against potential distraction or mental load attacks. Important objects could be modulated/highlighted in appearance (through AR), so that attention likelihood is above a certain threshold. In particular, virtual highlights for mission-critical physical objects are a promising direction for mitigation, given the better recall for virtual objects in studies 1 and 2. 
The system could strive for task simplification under high load, preserving object recall within operational thresholds. These approaches are plausible given our findings, but they will require additional experimental testing to evaluate their effectiveness.

Several limitations apply. Recall probes were not identical across studies. The datasets are from controlled AR search tasks with HoloLens 2, so generalization beyond similar conditions should be tested. While machine learning models are often benchmarked by transfer to new settings, SEM models differ in that they are designed to test theory-driven paths and highlight which factors consistently influence recall. The expectation is therefore not that this exact model applies unchanged to every future scenario, but that its structure can guide the inclusion of relevant indicators and be re-estimated with new data, even possibly utilizing additional statistical modeling approaches. In this way, SEM can provide continuity across studies while informing training and validation in each new environment.

In future iterations, we plan to add the remaining variables as seen in Figure \ref{fig:theorized_model}. Specifically, we believe there is valuable information in aggregated head rotation, distance traveled, and object density in the scenes. By adding in additional data points, not only will the model have more information, but we will also have more flexibility to improve the model design (such as including the latent variable Task).

\section{Conclusion}
Flares, drone clutter, decoys, and smoke can pull a Forward Observer’s attention off a mission-critical outpost. Our pooled PLS-SEM yields competitive $F_{1}$ and interpretable per-object recall probabilities that an ICS system can use as a control signal: when the predicted recall probability dips, suppress nonessential audio cues, reduce augmentation density, enforce stable world-locked guidance, and add a virtual outline or twin to increase target exposure time.

We model attention with an object-recall proxy in realistic mobile AR scenarios, compare PLS-SEM with Random Forests and an MLP under identical cross-validation, and show how the model can guide mitigations that help sustain attention during cognitive attacks. Next steps include adding impact factors such as head rotation, distance traveled, object density, and to test mitigation policies in FO and JTAC tasks.

\begin{acks}
This work was funded in part by the DARPA Intrinsic Cognitive Security Program under grant number \#HR001124C0409, via subcontract from SRI International. The underlying mobile AR studies were in part funded by ONR awards N00014-19-1-2553, N00174-19-1-0024, and N00014-23-1-2118. 
\end{acks}

\bibliographystyle{ACM-Reference-Format}
\bibliography{references,th_additions}

\end{document}